\documentclass{epl}
\usepackage{amsmath}

\title{Attraction and unbinding of like-charged rods \thanks{Revised version published in Europhys. Lett. {\bf 67}, 130 (2004).}}

\author{A. Naji\inst{1}\thanks{Present address: 
             Physics Dept., Technical University of Munich, 
             85748 Garching, Germany.} 
        \and A. Arnold\inst{2} 
        \and C. Holm\inst{2} 
        \and R. R. Netz\inst{1} ($^{\ast \ast}$)}
\shortauthor{A. Naji \etal}
\institute{
  \inst{1} Sektion Physik, 
           LMU, Theresienstr. 37, D-80333 M\"unchen, Germany. \\  
  \inst{2} 
           MPI f\"ur Polymerforschung, 
           Ackermannweg 10, D-55128 Mainz, Germany.}


\begin{document}
\maketitle
\vspace{-9mm}
\begin{abstract}
We investigate the effective interaction between two like-charged rods in the regime
of large coupling parameters using both Molecular Dynamics simulation techniques and 
the recently introduced strong-coupling theory. We obtain attraction between the rods 
for elevated  Manning parameters accompanied by an equilibrium surface-to-surface 
separation of the order of the Gouy-Chapman length. 
A continuous unbinding between the rods is predicted at a threshold Manning parameter
$\xi_c = 2/3$.
\end{abstract}

\vspace{-10mm}




\section{Introduction}
\label{sec:intro}

In recent years, both experiments \cite{Bloom} and numerical 
simulations \cite{Guld86,Gron97,Gron98,Wu,AllahyarovPRL,Linse,Stevens99,Messina,Deserno03} 
showed that like-charged macroions can attract each other 
via effective forces of electrostatic origin; one prominent example 
is the formation of dense packages of DNA molecules (DNA condensates) \cite{Bloom}. 
These observations indicate like-charge attraction in the presence of multivalent counterions,
at low temperatures or for large surface charge density on macroions, {\em i.e.}
when electrostatic correlations between charged particles become increasingly important. 
The standard mean-field Poisson-Boltzmann (PB) theory predicts 
repulsion between like-charged objects \cite{VO}.
In contrast,  incorporation of electrostatic correlations 
generates attractive interactions  in agreement with 
a variety of experimental and numerical results 
\cite{Kjellander84,Attard,Oosawa,Stevens90,Barbosa00,Rouzina96,Kornyshev,Shklo,Arenzon99,Netz01,Andre,Naji}.
The importance of electrostatic correlations can be quantified by means of
the {\em coupling parameter}, $\Xi=2\pi q^3 \ell_{\ab{B}}^2 \sigma_{\ab{s}}$, 
which depends on the 
charge valency of counterions, $q$,  
the surface charge density of macroions, $\sigma_{\ab{s}}$, 
and the Bjerrum length,
$\ell_{\ab{B}}=e^2/(4\pi \varepsilon\varepsilon_0 k_{\ab{B}} T)$
(associated with a medium of dielectric constant 
$\varepsilon$ and at temperature $T$). The PB theory 
is valid in the limit of vanishingly small coupling 
parameter $\Xi\rightarrow 0$
\cite{Netz01,Andre,Yoram04}, while non-mean-field features emerge 
in the opposite limit 
of large coupling, $\Xi\gg1$, and are typically 
accompanied by the formation of strongly-correlated counterion layers.
For charged {\em curved} surfaces, 
one has to consider the entropy-driven counterion-condensation 
process as well. 
For rod-like macroions, as considered in
this paper,  counterion condensation is controlled by 
the so-called {\em Manning parameter},
$\xi=q\ell_{\ab{B}} \tau$ \cite{Manning69}, 
where $\tau$ stands for the single-rod linear charge density
(in units of the elementary charge $e$).
For very  small Manning parameter, $\xi$, counterions de-condense from the rods
leading to a bare electrostatic repulsion between them \cite{Note00}. 
For sufficiently large $\xi$, on the other hand,  
a certain fraction of counterions remains bound to the rods 
and attraction becomes possible for moderate to large couplings.
We are interested in the regime of Manning parameters, $\xi$,
and coupling parameters, $\Xi$, 
where an effective rod-rod attraction is present.

Several  mechanisms for the  attraction between like-charged rods
have been considered in the literature, including
covalence-like binding \cite{Manning97}, 
Gaussian-fluctuation correlations \cite{Oosawa}, 
and structural correlations \cite{Kornyshev,Shklo,Arenzon99}. 
The two latter approaches  
yield attraction based on correlated fluctuations of condensed 
counterions on opposite rods and short-ranged correlations due to 
the ground-state configuration of the system, respectively. 
Conflicting predictions for the threshold  value of Manning parameter,
above which attraction between two rods is possible, 
have been obtained: The analysis of Ray and Manning \cite{Manning97}, 
based on the standard 
counterion-condensation model \cite{Manning69}, leads to 
attraction  for $\xi>1/2$. In contrast, 
an attraction regime of $\xi>2$ was proposed by the 
counterion-condensation theory of Arenzon {\em et al.} \cite{Arenzon99}. 
Recent numerical simulations \cite{Gron97,Deserno03}, 
on the other hand, reveal attraction 
already for Manning parameters of the order of  $\xi\approx 1$ and 
for moderate coupling parameters, though 
did not specifically consider the threshold value of $\xi$.
Recently, a systematic treatment of correlations in 
highly-coupled systems has been put forward \cite{Netz01}, 
that employs a perturbative 
scheme (virial expansion) in terms of $1/\Xi$.
This scheme leads to the asymptotic strong-coupling (SC) theory, 
which becomes exact in the 
limit which is complementary to the mean-field regime, 
{\em i.e.} when $\Xi\rightarrow \infty$. 
The SC theory has been used to study the interaction between planar 
charged walls \cite{Netz01}
and shows quantitative agreement with Monte-Carlo simulations for 
moderate to large coupling parameters \cite{Andre}. The SC mechanism
of like-charge attraction qualitatively agrees with the structural-correlation scenario 
\cite{Rouzina96,Kornyshev,Shklo,Arenzon99} 
for large macroion charges \cite{Netz01,Andre,Yoram04,Naji}.  
In this letter, we study the effective interaction in a system of 
two like-charged rods using both Molecular Dynamics (MD) 
simulation methods and the SC theory. 
The nature of electrostatic correlations in such a system has 
been studied in our previous 
MD simulations \cite{Deserno03}, and exhibits a competition  between 
electrostatic and excluded-volume interactions. 
Here, we focus on the strong-coupling 
characteristics of attraction between like-charged rods. In the following, 
these numerical results will be compared with the predictions 
of the SC theory \cite{Naji}.




\section{Strong-coupling theory}

In the limit of large coupling parameter $\Xi\gg 1$, 
the canonical free energy of a system of fixed 
macroions with their neutralizing counterions, ${\mathcal F}$, 
admits a large-coupling expansion
as ${\mathcal F}={\mathcal F}_1/\Xi+{\mathcal F}_2/\Xi^2+\ldots$ \cite{Naji}, 
where the coefficients ${\mathcal F}_1, {\mathcal F}_2,\ldots$ 
are expressed in terms of weighted integrals of Mayer functions over
counterionic degrees of freedom, which are convergent for systems of counterions at
charged macroscopic objects \cite{Netz01}. The first term in the above expression
generates the leading non-vanishing contribution to statistical observables such as 
effective interaction between macroions for $\Xi\gg 1$;
hence ${\mathcal F}_{\ab{SC}}={\mathcal F}_1/\Xi$ is referred to as the SC free energy
of the system and  ${\mathcal F}_1$ as the rescaled SC free energy. 
The SC free energy takes a very simple form in terms of the one-particle interaction
energies between counterions and macroions 
(see, {\em e.g.}, eq.~(\ref{eq:SCfree}) below), and the
mutual interaction between counterions themselves enters only in the sub-leading 
corrections \cite{Netz01,Naji}. Physically, this reflects the fact that for $\Xi\rightarrow \infty$, 
counterions at charged objects are surrounded by an extremely large correlation 
hole and thus, the single-particle term becomes the dominant 
contribution \cite{Netz01,Andre,Yoram04,Naji} (see also the discussion).  
\begin{figure}[t]
  \onefigure[angle=0.0,scale=0.335]{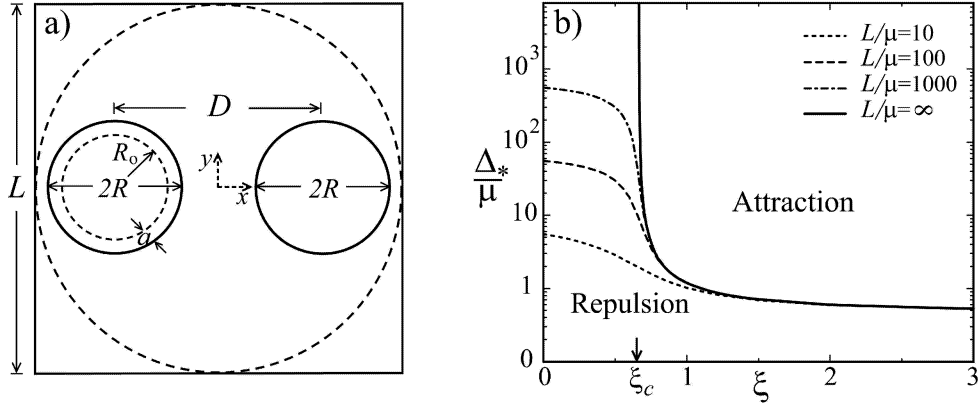}
  \caption{a) We consider two identical rods of radius $R$ at 
    axial distance $D$ in an outer
    box of size $L$, which is rectangular (with square cross-section) 
     in the analytical model and cylindrical in the simulations. 
    $R_0$ and $a$ are the Lennard-Jones potential parameters defined 
    in the simulation model. 
    b) The rescaled equilibrium surface-to-surface distance of rods, 
    ${\tilde \Delta}_\ast=\Delta_\ast/\mu=(D_\ast- 2R) / \mu$, obtained from the SC theory 
    as a function of Manning parameter, $\xi=q\ell_{\ab{B}} \tau$, for various rescaled
    box sizes, $L/\mu$, indicated on the graph. In the infinite-box limit, 
    a continuous unbinding occurs at $\xi_c  = 2/3$. }
  \label{fig:fig1}
 \vspace{-5mm}
\end{figure}
Let us consider two infinitely long similar rods of length $H$ and radius $R$ that are 
located parallel at axial separation $D$ in a rectangular box of 
lateral extension $L$ containing also neutralizing counterions
of valency $q$ (fig.~\ref{fig:fig1}a). 
The rods are impenetrable to counterions and no additional salt is present.
It is convenient to establish a dimensionless formalism for such
systems by rescaling all length scales by the Gouy-Chapman length 
$\mu=1/(2\pi q \ell_{\ab{B}} \sigma_{\ab{s}})$, such as ${\tilde x}=x/\mu$, {\em etc.} 
\cite{Netz01}, where $\sigma_{\ab{s}}=\tau/(2\pi R)$ stands for the surface 
charge density of the rods. Clearly, the rod radius in rescaled units, ${\tilde R}=R/\mu$, 
coincides with Manning parameter, {\em i.e.} ${\tilde R}=\xi=q \ell_{\ab{B}} \tau$.
(We may occasionally use $\xi$ or ${\tilde R}$ to denote the Manning parameter.)
The rescaled rod length, ${\tilde H}$, is related to the total number of counterions, $N$, 
via the global electroneutrality condition as ${\tilde H}=N \Xi/\xi$. 
The rescaled SC free energy of this system per rescaled unit length follows as \cite{Naji}
\begin{equation}
   \frac{{\mathcal F}_1}{\tilde H}= 
                  -2 \xi^2\ln {\tilde D}
                  -2\xi \ln\left[\int_V \upd{\tilde x}\upd{\tilde y}\,\,
                       {\tilde r}_1^{-2\xi}{\tilde r}_2^{-2\xi}\right].
\label{eq:SCfree}
\end{equation} 
The integral runs over the volume $V$ available for counterions within the confining box, 
excluding the rods. We defined  
${\tilde r}_{1,2}=[({\tilde x}\pm{\tilde D}/2)^2+{\tilde y}^2]^{1/2}$ 
as the radial distances from the rods axes. The first term in 
eq.~(\ref{eq:SCfree}) is nothing but the bare electrostatic repulsion
between the rods. The
second term involves the leading (energetic and entropic) contributions
from counterions, which leads to a counterion-mediated attraction between sufficiently 
highly charged rods \cite{Note0}. This term also  
reflects the de-condensation process of counterions 
at low Manning parameters: For $\xi<1/2$, the counterionic integral 
in eq.~(\ref{eq:SCfree}) diverges in the infinite-volume limit 
${\tilde L}\rightarrow \infty$, hence the distribution of 
counterions around the rods and the counterion-mediated force 
vanish, {\em i.e.} the rods purely repel each other \cite{Note00}.  
For large Manning parameter $\xi\gg 1$, on the other hand, 
the rescaled SC free energy, ${\mathcal F}_1$, 
shows a long-ranged attraction and a pronounced global
minimum at a small axial separation
${\tilde D}_\ast\approx 2{\tilde R}$, which is nothing but
the  equilibrium axial separation of the rods \cite{Naji}. 
For $\xi\gg 1$, the counterionic distribution becomes strongly localized in a narrow region 
between the rods, indicating that the SC attraction is accompanied by strong 
accumulation of counterions in the intervening region. 
This  allows for a saddle-point calculation of 
the counterionic integral in eq.~(\ref{eq:SCfree}), giving  the approximate form of 
${\mathcal F}_1$ in the vicinity of its local minimum as \cite{Naji}
\begin{equation} 
  {\mathcal F}_1/{\tilde H} \approx 6 \xi^2\ln {\tilde D}
              -2 \xi \ln ({\tilde D}-2{\tilde R}).
\label{eq:SCfree_Rlarge}
\end{equation}  
The  equilibrium surface-to-surface distance of the rods for large $\xi$
is  obtained by minimizing expression (\ref{eq:SCfree_Rlarge}) as 
${\tilde \Delta}_\ast\equiv {\tilde D}_\ast-2{\tilde R}\approx 2/3+{\mathcal O}(1/\xi)$, 
which corresponds, in real units, to a surface separation of the order of the 
Gouy-Chapman length, $\mu$ \cite{Note2}. Figure~\ref{fig:fig1}b shows the 
global behavior of the rescaled equilibrium surface-to-surface distance, 
${\tilde \Delta}_\ast$, obtained  by numerical minimization of
the full SC free energy, eq.~(\ref{eq:SCfree}), as a function of $\xi$ 
for several box sizes. 
For large $\xi$, attraction is dominant, the rods form a closely-packed bound state,
and the volume of the bounding box
is irrelevant  due to strong condensation of counterions. However,  for 
decreasing $\xi$, attraction is 
weakened and rods eventually unbind in a continuous fashion
when ${\tilde L}\rightarrow \infty$. For
two unconfined rods (solid curve in fig.~\ref{fig:fig1}b), the unbinding transition occurs
at $\xi_c=2/3$ and exhibits an asymptotic power-law behavior as a function
of the  reduced Manning parameter as 
${\tilde \Delta}_\ast\sim (\xi -2/3)^{-\alpha}$, where $\alpha\approx 3/2$ \cite{Naji}.
Note that the predicted onset of rod-rod attraction, $\xi_c=2/3$, is
somewhat larger than the 
counterion-condensation threshold for the combined system
of two rods, $\xi=1/2$ \cite{Manning97}, and smaller than the condensation threshold for 
a single rod, $\xi =1$ \cite{Manning69}. This reflects the subtle crossover from 
the case of two coupled rods (when the distance between them is small) to 
the case of two decoupled rods as the axial distance diverges when the 
unbinding threshold is approached.
The predicted attractive force between rods,
$F_{\ab{rods}}= -(k_{\ab{B}}T)\partial {\mathcal F}_{\ab{SC}}/\partial D$ (in actual units),
is found from eq.~(\ref{eq:SCfree_Rlarge}) to be inversely proportional 
to the axial distance, $D$. This force increases and tends to a temperature-independent value 
upon increasing $\xi$ (or decreasing temperature) \cite{Gron97}, which is nothing but the 
{\em energetic attraction} mediated by counterions sandwiched between closely-packed rods, {\em i.e.}
$F_{\ab{rods}}=-3e^2\tau^2/(2\pi \varepsilon \varepsilon_0 D)$ \cite{Naji}. The SC attraction
for large $\xi$, thus, originates from the energetic contributions induced by
the ground-state configuration of counterions and in this respect,
agrees qualitatively with the low-temperature picture for like-charge 
attraction \cite{Rouzina96,Kornyshev,Shklo,Arenzon99}. Though for the specific case 
of two charged rods, the quantitative predictions of the SC theory for the (energetic) 
inter-rod force at large $\xi$, and for the attraction threshold differ 
from the results of the model studied in ref.~\cite{Arenzon99},
which incorporates a discrete charge pattern for the rods and obtains attraction 
for $\xi>2$ \cite{Arenzon99}.




\section{Simulation method}

\begin{figure}[t]
\onefigure[angle=0,scale=0.335]{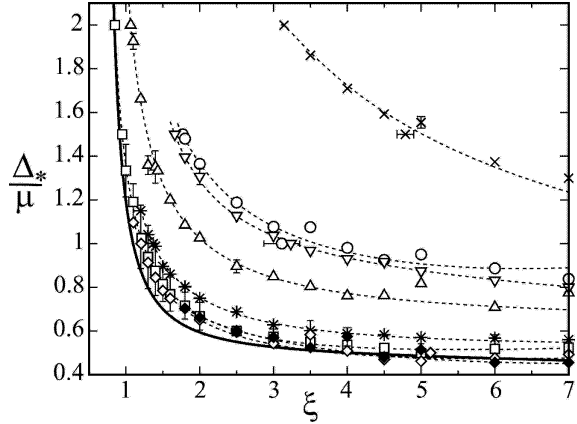}
\caption{The equilibrium surface-to-surface distance of two rods 
in rescaled units, ${\tilde \Delta}_\ast=\Delta_\ast/\mu=(D_\ast-2R)/\mu$,
as a function of Manning parameter, $\xi=q \ell_{\ab{B}} \tau$. 
Symbols show simulation results 
for (from top): $\gamma_{\ab{RB}}=3$ (crosses), 9.1 (open circles), 
10 (open triangle-downs), 15 (open triangle-ups), 30
(stars), 40 (open squares), 50 (open diamonds), 60 (filled diamonds). The solid curve 
is the SC prediction obtained by 
numerical minimization of eq.~(\ref{eq:SCfree}).  
Dashed curves are guides to the eye.} 
\label{fig:fig2}
 \vspace{-5mm}
\end{figure}

To study the interaction between two like-charged rods numerically, we have 
performed extensive Molecular Dynamics 
simulations making use of a Langevin thermostat to drive the 
system into the canonical state. 
The geometry of the simulated system is similar to what 
we have sketched in fig.~\ref{fig:fig1}a;  the outer confining box 
is here chosen to be cylindrical. Other parameters are defined 
in the same way unless explicitly mentioned.
Charged particles interact via bare Coulombic interaction and
excluded-volume interactions. We apply periodic boundary conditions 
in the direction parallel to the rods axes. This 
leads to summation of infinite series for Coulomb interaction over all periodic 
images, which is handled numerically using the MMM1D summation method \cite{mmm2d}. 
As we are only interested in examining electrostatic aspects of the 
effective rod-rod interaction, excluded-volume interactions are considered
only between rods and counterions and not between  counterions
themselves. (Extended numerical results with excluded-volume interaction between 
counterions will be presented elsewhere \cite{Arnold03}.)
We employ a shifted Lennard-Jones potential as
\begin{equation}
  V_{\ab{LJ}}(r) = 
  \begin{cases}
    4\,\epsilon
    \left[\left(\frac{a}{r-R_0}\right)^{12} -
      \left(\frac{a}{r-R_0}\right)^6 +
      \frac{1}{4} \right] &\quad : \quad (r-R_0) < 2^{1/6}a \\
    0  &\quad : \quad \text{otherwise}
  \end{cases} ,
\label{eq:LJ}
\end{equation}
where $R_0$ is an offset used to control the radius of the rods,  
$a$ defines our basic length scale in the simulations, and 
$\epsilon\approx 1 k_{\ab{B}}T$.   
Therefore, $R=R_0+a$ may be regarded as the effective rod radius and
is directly compared with the SC result (with hard-core rods). 
This is justified since in the simulations only a negligible 
fraction of counterions penetrates the above potential to reach radial 
distances smaller than $R$. 
In the simulations, the diameter of the (cylindrical) outer box is chosen as $8D$. 
For the final comparison (fig.~\ref{fig:fig2}), the theoretical curves are also
calculated using a similar constraint, though for simplicity and as explained before, 
calculations have been done for a square box of edge size $L=8D$. 
As we have explicitly checked, the results are insensitive to small changes in the 
box size for the considered range of Manning parameter 
(see also fig.~\ref{fig:fig1}b), and the SC predictions
can thus be compared with the simulations.
To obtain the location of the zero-force (equilibrium) point 
in the force curves, we have employed
a bisection algorithm followed by a linear regression. 
The error bars are determined from the error of the 
linear regression using the error propagation method.

\section{Results and discussion}

To clearly separate the attraction and repulsion regimes of two like-charged rods numerically,
we study the rescaled equilibrium surface-to-surface separation of rods, ${\tilde \Delta}_\ast$, 
as a function of Manning parameter, $\xi$, in our simulations. The results will then be compared with
the SC prediction (fig.~\ref{fig:fig1}b). 
To this end, we fix the actual surface-to-surface distance between the rods, 
$\Delta=D-2R$, and their linear charge density, $\tau$, together with the counterion valency,
$q$, but vary the Bjerrum length, $\ell_{\ab{B}}$, and the actual rod radius, $R$. 
By doing so, the Gouy-Chapman length associated with this system, 
$\mu=R/(\ell_{\ab{B}} q \tau)$, is varied, which then allows for determining 
the equilibrium surface-to-surface separation {\em in rescaled units}, 
${\tilde \Delta}_\ast=\Delta_\ast/\mu$.
The results will also depend upon the electrostatic coupling parameter,
$\Xi=2\pi q^3 \ell_{\ab{B}}^2 \sigma_{\ab{s}}$, which is finite in the simulations
and may be written as 
\begin{equation}
    \Xi=\xi\,{\tilde \Delta}\,\gamma_{\ab{RB}},
\label{eq:coupling}
\end{equation}
where $\gamma_{\ab{RB}}=q/(\tau \Delta)$
is a dimensionless parameter referred to here as the
{\em Rouzina-Bloomfield parameter} (see the discussion below). 
The simulation results for the equilibrium 
surface-to-surface separation of the rods 
are shown in fig.~\ref{fig:fig2} for different  
$\gamma_{\ab{RB}}$ ranging from 3 to 60.
The data have been obtained using various combinations of fixed
$\tau a=1.0, 0.33, 0.10$,
$q=1, 3, 5, 10$ and $\Delta/a=0.5, 1.0, 2.0, 2.5$.
For a given set of system parameters ($\tau$, $q$ and $\Delta$) chosen in the
simulations, $\gamma_{\ab{RB}}$ is fixed and identifies a single curve. 
As seen, upon increasing the Rouzina-Bloomfield parameter, $\gamma_{\ab{RB}}$, 
the equilibrium separation decreases indicative of a stronger attraction operating 
between the rods, and at the same time, the agreement between simulation data and the 
SC prediction  (solid curve in fig.~\ref{fig:fig2})
becomes progressively better. The agreement 
is quantitative for large $\gamma_{\ab{RB}}$, {\em i.e.}
$\gamma_{\ab{RB}}=50$ and 60. (The data set with $\gamma_{\ab{RB}}=60$ is obtained using 
$q=3$, $\tau a=0.1$ and $\Delta/a=0.5$, and data with 
$\gamma_{\ab{RB}}=50$ are obtained using $q=5$, $\tau a=0.1$ and $\Delta/a=1.0$.) 
The observed trend for increasing $\gamma_{\ab{RB}}$ is associated 
with the increase of the coupling parameter, $\Xi$, in the simulated 
system (see eq.~(\ref{eq:coupling})), which eventually exhibits 
the strong-coupling regime. (For instance, for a moderate Manning parameter
of $\xi=3.0$, $\Xi$ increases from 18 for cross symbols up to about 100 
for filled diamonds.)

It is noteworthy that $\gamma_{\ab{RB}}$, as defined above, can be expressed 
as $\gamma_{\ab{RB}}=\delta/\Delta$, where $\delta=q/\tau$ is the projected
distance between counterions along a single rod as follows from the 
local electroneutrality condition. Note that $\delta$ roughly gives the 
correlation hole size around counterions at charged 
surfaces \cite{AllahyarovPRL,Linse,Deserno03,Andre,Yoram04}. 
The appearance of strong attractive forces for $\gamma_{\ab{RB}}>1$ 
was first addressed by Rouzina and Bloomfield 
\cite{Rouzina96} for a system of two planar charged walls. 
Analytical and numerical results for planar walls \cite{Netz01,Andre,Yoram04} have in fact
shown that the {\em relative} magnitude of higher-corrections to the asymptotic SC theory
decreases with $\gamma_{\ab{RB}}$, and the SC regime 
at finite $\Xi$ is characterized by the Rouzina-Bloomfield criterion $\gamma_{\ab{RB}}>1$.
Physically, this corresponds to a large correlation hole size around counterions 
at macroion surfaces as compared with the macroions surface separation, 
{\em i.e.} $\delta>\Delta$, leading to a dominant contribution from single (isolated)
counterions \cite{Netz01,Andre,Yoram04}, 
which is formally obtained within the SC scheme \cite{Netz01}.
It becomes exceedingly difficult to perform systematic analytic calculations to examine 
higher-order corrections to SC predictions 
given the geometry of the two-rod system. Nonetheless, the 
present numerical results clearly indicate a qualitatively similar trend for 
increasing $\gamma_{\ab{RB}}$ in this system and that, the SC regime 
is characterized by Rouzina-Bloomfield criterion. 
Due to convergence limitations, our numerical investigation so far has been 
limited to the range of $\xi>0.8$ and thus do not cover the 
close vicinity of the unbinding transition. (It becomes more difficult to obtain good data as the 
rods equilibrium distance rapidly increases for small $\xi$.) However, the excellent
convergence of the simulation results to the asymptotic SC
prediction suggests an attraction threshold of about $\xi_c = 2/3$ 
for two rods in the strong-coupling limit as obtained using the SC theory.
Note that for small Manning parameters where counterions de-condense ($\xi\leq 1/2$), 
electrostatic correlations are suppressed due to entropic dilution of the counterionic 
cloud. Physically, this regime does not exhibit strong energetic coupling and higher-order
corrections to the SC theory may become important.
Nevertheless as we showed, the SC theory captures 
the de-condensation process and leads to a qualitatively consistent picture for the whole range of 
Manning parameters \cite{Note00}. (Note that the predicted onset of de-condensation at $\xi=1/2$ and also 
the bare macroionic repulsion for small $\xi$ {\em quantitatively} agree with 
previous findings \cite{Manning97}.)   
How and whether the predicted attraction threshold ($\xi_c = 2/3$) and the unbinding behavior
is influenced by the de-condensation process of counterions at small $\xi$ remains to be clarified.

In summary, we have investigated the attraction and repulsion regimes of two 
like-charged rods in terms of Manning parameter, $\xi$, and the coupling 
parameter, $\Xi$,
by combining numerical and analytical approaches. For large $\xi$,
the rods form a closely-packed bound state of small surface-to-surface separation 
of about the Gouy-Chapman length, $\mu$ \cite{Note2}. The attraction is weakened and the 
rods unbind for decreasing $\xi$. 
The attraction threshold is found at $\xi_c=2/3$, and   
the unbinding of rods proceeds in a continuous fashion characterized by 
a power-law  for the rods surface separation, 
$\Delta_\ast\sim (\xi-2/3)^{-\alpha}$, with an exponent $\alpha\approx 3/2$.
The predicted attraction regime for large $\xi$ is accessible 
experimentally using, {\em e.g.}, 
DNA molecules ($\tau\, e\approx 6\,e/\un{nm}$) in the presence of
multivalent counterions such as spermidine ($q=3$), which yields 
$\xi\approx 12$ and $\Xi\approx 80$. For these values, our results predict attraction 
with an equilibrium surface separation $\Delta_\ast\sim \mu$, 
where for the DNA-spermidine system $\mu\sim 1$\AA \cite{Note2}. 
It should be noted, however, that for DNA-like molecules (large $\tau$), other factors
such as excluded-volume interaction between counterions \cite{Deserno03,Arnold03} and the
helical structure of DNA \cite{Kornyshev,Shklo} may become important and contribute 
additional components to the total effective force. 
The predicted continuous unbinding occurs at quite small values of Manning parameter.
Experimentally, this phenomenon could be studied with weakly charged stiff polymers,
such as poly-phenylene with a suitably chosen small density of charged side-chains.
The effects of salt can be qualitatively accounted 
for by associating the bounding box size
with the screening length: Thus only close to the unbinding do 
we expect addition of salt to matter.
The effects of finite chain stiffness and the bundling of many 
chains constitute interesting applications for the future.
\vspace{-3mm}

\acknowledgments 
A.N. and R.R.N. acknowledge funds from DFG German-French Network.  
C.H. and A.A. acknowledge funds from  the Zentrum f\"ur
Multifunktionelle Werkstoffe und Miniaturisierte Funktionseinheiten,
grant BMBF 03N 6500, DFG grant Ho 1108/11-1, and SFB 625.
\vspace{-3mm}


\end{document}